\documentclass[preprint,showpacs,preprintnumbers,amsmath,amssymb]{revtex4}

\usepackage{graphicx}

\newcommand{\AddrLNF}{
  {\it INFN, Laboratori Nazionali di Frascati,C.P. 13, I00044 
    Frascati, Italy}}
\newcommand{\AddrIfTP}{
  {\it Institute for Theoretical Physics, Kanazawa University, 
    920-1192 Kanazawa, Japan}}
\newcommand{\AddrUdeA}{
  {\it Instituto de F\'\i sica, Universidad de Antioquia,
    A.A.{\it{1226}}, Medell\'\i n, Colombia}}
\newcommand{\AddrEIA}{
  {\it Escuela de Ingenier\'{\i}a de Antioquia,
    Calle 25 sur No 42-73, Envigado, Colombia}}
\begin{document}
\preprint{arXiv:0808.3340}

\title{Radiative seesaw model: Warm dark matter, collider and lepton
  flavour violating signals}

\author{D. Aristizabal Sierra}
\affiliation{\AddrLNF}
\author{Jisuke Kubo}
\author{Daijiro Suematsu}
\affiliation{\AddrIfTP}
\author{D. Restrepo}
\author{Oscar Zapata}
\altaffiliation[Also at]{\AddrEIA}
\affiliation{\AddrUdeA}

\date{\today}
\begin{abstract}
  Extending the standard model with three right-handed neutrinos
  ($N_k$) and a second Higgs doublet ($\eta$), odd under the discrete
  parity symmetry $Z_2$, Majorana neutrino masses can be generated at
  1-loop order. In the resulting model, the lightest stable particle,
  either a boson or a fermion, might be a dark matter candidate. Here
  we assume a specific mass spectrum ($M_1\ll M_2 < M_3 < m_\eta$) and
  derive its consequences for dark matter and collider phenomenology.
  We show that (i) the lightest right-handed neutrino is a warm dark
  matter particle that can give a $\sim$10\% contribution to the dark
  matter density; (ii) several decay branching ratios of the charged
  scalar can be predicted from measured neutrino data. Especially
  interesting is that large lepton flavour violating rates in muon and
  tau final states are expected. Finally, we derive upper
  bounds on the right-handed neutrino Yukawa couplings from the current
  experimental limit on $Br(\mu\to e\gamma)$.
\end{abstract}

\pacs{12.60.Fr,13.15.+g,14.60.Pq,14.60.St}

\maketitle
\section{Introduction}
\label{sec:int}
Solar~\cite{Ahmad:2002jz}, atmospheric~\cite{Fukuda:1998mi} and
reactor~\cite{Eguchi:2002dm} neutrino experiments have demonstrated
that neutrinos have mass and non-zero mixing angles among the
different generations. On the other hand observations of the cosmic
microwave background, primordial abundances of light elements and
large scale structure formation have firmly established that most of
the mass of the Universe consists of dark matter
(DM)~\cite{Spergel:2006hy}. These experimental results are at present
the most important evidences for physics beyond the standard model.

There are several ways in which neutrino masses can be
generated. Certainly the best-known mechanism to generate small
Majorana neutrino masses is the seesaw~\cite{Minkowski:1977sc}.
However, a large variety of models exist in which lepton number is
broken near-or at- the electroweak scale. Examples are supersymmetric
models with explicit or spontaneous breaking of
R-parity~\cite{Hirsch:2004he,Hirsch:2008ur}, models with Higgs
triplets~\cite{Garayoa:2007fw}, pure radiative models at
1-loop~\cite{Zee:1980ai} or at
2-loop~\cite{Zee:1985id} order and models in
which neutrino masses are induced by leptoquark
interactions~\cite{AristizabalSierra:2007nf}.

According to their free-streaming length DM particle candidates can be
classified as either hot, warm or cold DM. Due to their large
free-streaming length the mass and density of hot DM particles are
strongly constrained~\cite{Hannestad:2003ye}\footnote{Models where all
dark matter is hot are ruled out completely by current cosmological
data~\cite{Hannestad:2005fg}}. Contrary, cold DM particles have a
free-streaming length which is irrelevant for cosmological structure
formation. Actually, cold DM is usually considered the best choice to
fit large scale structure data~\cite{Primack:1997av}. Warm DM (WDM)
particles, for instance those that decouple very early from the
thermal background, have a smaller temperature than that of hot dark
matter relics and thus a shorter free-streaming length.

It has been argued in the literature~\cite{Bode:2000gq} that WDM
scenarios may be able to overcome the shortcomings of the standard
cold DM scenario.  Constraints on WDM particles have been quoted in
reference~\cite{Viel:2005qj}. If DM consists only of WDM,
$m_{\mbox{\tiny{WDM}}}\gtrsim 1.2$  keV 
whereas in mixed scenarios, in
which the DM relic density receives contributions from cold and WDM as
well, $m_{\mbox{\tiny{WDM}}}\lesssim 16$ eV\footnote{In this case WDM
gives a contribution of $\sim$10\% to the total DM relic
density~\cite{Viel:2005qj}}.

The question of whether neutrino mass generation and DM are related
has lead to a large number of models~\cite{Kubo:2006rm}. In this paper
we focus on a particular realization, namely the radiative seesaw
model~\cite{Ma:2006km}.  In this scheme three right-handed neutrinos
$N_i$ and a second Higgs doublet $\eta=(\eta^+, \eta^0)$, odd under
the discrete parity symmetry $Z_2$, are added to the standard model.
As a result $(a)$ the new Higgs doublet has a zero vacuum expectation
value and there is no Dirac mass term. Thus, neutrinos remain massless
at tree level; $(b)$ the lightest particle in the spectrum, either a
boson or a fermion, is stable and therefore, in principle, can be a
dark matter candidate~\cite{Kubo:2006yx}.

Here we study the implications for DM and possible collider signatures
of this model. Our analysis is done in a particular scenario in which
the Yukawa couplings of $N_3$ are larger than those from $N_2$ and the
right-handed neutrino spectrum is such that $M_1\ll M_2< M_3$.
The right-handed neutrinos are assumed to be always lighter than the
charged and neutral scalars. As it will be shown the lightest neutrino
singlet can not be a cold DM candidate and instead behaves as WDM,
contributing with less than 10\% to the total DM relic density. In
addition we will show that current experimental neutrino data enforces
a number of constraints on the parameter space of the model. These
constraints, in turn, can be used to predict the decay patterns of the
charged scalar $\eta^\pm$. Therefore, the hypothesis that this model
is responsible for the generation of neutrino masses (within our
scenario) and that $N_1$ is a WDM particle can be tested in collider
experiments.

The rest of this paper is organized as follows: in section
\ref{sec:neutrinomassgen} we briefly describe the model, paying special
attention to the neutrino mass generation mechanism. In
section~\ref{sec:analytical-results} we present simple and useful
analytical results for neutrino masses and mixing angles.  In
section~\ref{sec:scalarDM} we discuss dark matter within the model and
show that the lightest right-handed neutrino is a WDM relic. We then
turn to the collider phenomenology of charged scalars in
section~\ref{sec:coll-sig}.  We show that different ratios of
branching ratios of $\eta^\pm$ can be predicted from measured neutrino
mixing angles. In section \ref{sec:fvclepdecays} we analyse the
implications of the model for lepton flavour violating decays, in
particular for $\mu\to e\gamma$. Finally in section~\ref{sec:summary}
we present our conclusions.
\section{Neutrino mass generation}
\label{sec:neutrinomassgen}
The model we consider~\cite{Ma:2006km} is a simple extension of the
standard model, containing three $SU(2)_L\times U(1)_Y$ fermionic
singlets $N_i$ and a second Higgs doublet $\eta$. In addition, an
exact $Z_2$ discrete symmetry is assumed such that the new fields are
odd under $Z_2$ whereas the standard model fields are even. The Yukawa
interactions induced by the new Higgs doublet are given by
\begin{equation}
  \label{eq:yukint}
  {\cal  L}=\epsilon_{ab}
  h_{\alpha j}\overline{N}_j P_L L_\alpha^a \eta^b + \mbox{h.c.}
\end{equation}
Here, $L$ are the left-handed lepton doublets, $\alpha, j$ are
generation indices (Greek indices label lepton flavour $e, \mu, \tau$) and
$\epsilon_{ab}$ is the completely antisymmetric tensor.  Apart from
these Yukawa interactions the quartic scalar term
\begin{equation}
  \label{eq:lambda5}
  \frac{1}{2}\lambda_5(\phi\eta)^2\,,
\end{equation}
where $\phi$ is the standard model Higgs doublet, is also relevant for 
neutrino mass generation. Since $Z_2$ is assumed to be an exact symmetry 
of the model $\eta$ has zero vacuum expectation value. Thus, there is no 
mixing between the neutral CP-even (CP-odd) components of the Higgs 
doublets. The physical scalar bosons are, therefore,
$\mbox{Re}\,\phi^0, \eta^\pm, \eta^0_R\equiv\mbox{Re}\,\eta^0$ and
$\eta^0_I\equiv\mbox{Im}\,\eta^0$.

\begin{figure}[t]
  \centering
  \includegraphics{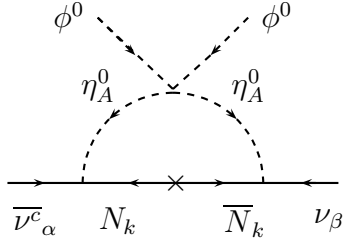}
  \caption{Feynman diagram for Majorana neutrino masses. $A=R,I$ labels
    the contributions from the neutral CP-even and CP-odd components
    of the Higgs doublet $\eta$.}
  \label{fig:neugen}
\end{figure}
The setup of equation (\ref{eq:yukint}) and equation (\ref{eq:lambda5})
generates Majorana neutrino masses through the diagram shown in figure
\ref{fig:neugen}. The resulting neutrino mass matrix can be written
as
\begin{equation}
  \label{eq:nmmexact}
  ({\cal  M_\nu})_{\alpha\beta}=\frac{1}{16\pi^2}\sum_{\substack{A=I,R\\k=1\dots 3}}
  c_A\,M_k\,h_{\alpha k}\,h_{\beta k}\,B_0(0,m_A^2,M_k^2) \,.
\end{equation}
Here $A=R,I$, $M_k$ are the right-handed neutrino masses, $m_A$ are the
$\eta_A^0$ masses, $c_R=+1$ while $c_{I}=-1$ and $B_0(0,m_A^2,M_k^2)$ is a 
Passarino-Veltman function \cite{Passarino:1978jh}. The function $B_0$
has a finite and an infinite part. Note that the infinite part cancel 
after summing over $A$ and the resulting formula can be expressed as a 
{\it difference} of two $B_0$ functions. The finite part of the 
Passarino-Veltman function $B^f_0$ is given by
\begin{equation}
  \label{eq:pass-velt}
  B_0^f(0,m_A^2,M_k^2)=\frac{m_A^2\log(m_A^2)-M_k^2\log(M_k^2)}
  {m_A^2-M_k^2}\,.
\end{equation}

As pointed out in reference\cite{Ma:2006km} if $\eta_R$ and $\eta_I$
are almost degenerate, i.e. $m_R^2-m_I^2=2 \lambda_5 v^2$ 
($v^2=(2\sqrt{2}G_F)^{-1}$) is assumed to be small compared to 
$m^2_0=(m_R^2+m_I^2)/2$, the neutrino mass matrix in (\ref{eq:nmmexact}) 
can be rewritten as
\begin{equation}
  \label{eq:nmmappx}
  ({\cal  M_\nu})_{\alpha\beta}=\frac{\lambda_5\,v^2}{8\pi^2}\sum_{k=1\dots 3}
  \frac{h_{\alpha k}\,h_{\beta k}M_k}{m_0^2-M_k^2}
  \left[
    1 - \frac{M_k^2}{m_0^2-M_k^2}\,
    \log
    \left(
      \frac{m_0^2}{M_k^2}
    \right)
  \right]\,.
\end{equation}
Depending on the relative size between $m_0$ and $M_k$ this formula
can be simplified~\cite{Ma:2006km}. Here we will focus on the limiting case
$m_0^2\gg M_k^2$.
\section{Analytical  results}
\label{sec:analytical-results}
Here we will consider a right-handed neutrino spectrum such that
$M_1\ll M_2< M_3$. In addition, as previously mentioned, we will also
consider the limiting case $m_0^2\gg M_k^2$. In this case the neutrino 
mass matrix in eq.~(\ref{eq:nmmappx}) becomes
\begin{equation}
  \label{eq:nmmfermDMcase}
  ({\cal  M_\nu})_{\alpha\beta}=\frac{\lambda_5\,v^2}{8\pi^2m_0^2}\sum_{k=1\dots 3}
  h_{\alpha k}\,h_{\beta k}M_k \,.
\end{equation}
In general the neutrino mass matrix receives contributions from
diagrams involving the three right-handed neutrinos. However, if $N_1$
is light enough, let us say, ${\cal O}(M_1/M_2)< 10^{-2}$ the
contributions from $N_1$ become negligible. In this limit
$\mbox{Det}[{\cal M}_\nu]\simeq 0$ and therefore only two neutrinos
have non-zero masses. In this case simple analytical formulas
involving neutrino mixing angles and Yukawa couplings can be derived.
Note that in this limit only a hierarchical spectrum is possible.  In
what follows we will focus on the normal spectrum. Some comments on
the inverted one will be given in section
\ref{sec:collider-signals-neutrino-pp}.

In the limit $\mbox{Det}[{\cal  M}_\nu]\simeq 0$ the mass matrix structure
is determined by the Yukawa couplings $h_{\alpha(2,3)}$. Therefore, it is useful 
to define two vectors in parameter space
\begin{eqnarray}
  \label{eq:fermionDMvectors}
  \mathbf{h_2}=(h_{12},h_{22},h_{32})\,,\nonumber\\
  \mathbf{h_3}=(h_{13},h_{23},h_{33})\,.
\end{eqnarray}
In terms of these vectors the two non-zero neutrino masses can be written 
as
\begin{equation}
  \label{eq:fermionicDMeig}
  m_{\nu_{2,3}}={\cal  G}_f
  \left[
    1
    \mp
    \sqrt{1- 
      4r_N\frac{
      |\mathbf{h_2}|^2|\mathbf{h_3}|^2
      -
      |\mathbf{h_2}\cdot \mathbf{h_3}|^2
      }{(r_N|\mathbf{h_2}|^2 + |\mathbf{h_3}|^2)^2}
    }
  \right]\,,
\end{equation}
where ${\cal  G}_f$ is given by
\begin{equation}
  \label{eq:globalfactor}
  {\cal  G}_f=\frac{\lambda_5 v^2 M_3}{16 \pi^2 m_0^2}
  \,(r_N|\mathbf{h_2}|^2 + |\mathbf{h_3}|^2)
\end{equation}
and
\begin{equation}
  \label{eq:r-h-n-ratio}
  r_N=\frac{M_2}{M_3}\,.
\end{equation}
The ratio between the solar and the atmospheric scale 
is approximately given by
\begin{equation}
  \label{eq:atmospsolarR}
  R\equiv\sqrt{\frac{\Delta m^2_{21}}{\Delta m^2_{32}}}\simeq
  \frac{m_{\nu_2}}{m_{\nu_3}}\,.
\end{equation}
Thus, from eq.~(\ref{eq:fermionicDMeig}) and (\ref{eq:globalfactor}),
it can be noted that $R$ is independent of ${\cal  G}_f$ and therefore
independent of $\lambda_5$ and $m_0$. 

The generation of the non-zero lightest neutrino mass can be understood
from the misalignment angle
between the parameter space vectors $\mathbf{h_{2,3}}$
($\cos\theta=\mathbf{h_2}\cdot\mathbf{h_3}/|\mathbf{h_2}||\mathbf{h_3}|$)
which, from eq.~(\ref{eq:fermionicDMeig}), can be written as
\begin{equation}
  \label{eq:cosSqtheta}
  \sin^2\theta =
  \frac{(1 + h_r \,r_N)^2}{4 h_r\, r_N}
  \left[
    1 - \left(\frac{1-R}{1+R}\right)^2
  \right]\,,
\end{equation}
where $h_r=|\mathbf{h_2}|^2/|\mathbf{h_3}|^2$. Note that since $h_r$
as well as $r_N$ are positive quantities a complete alignment between
$\mathbf{h_2}$ and $\mathbf{h_3}$ ($\sin\theta=0$) is only possible
if $R=0$. However, this possibility is excluded as it implies 
$m_{\nu_2}=0$.

\begin{figure}[t]
  \centering
  \includegraphics[width=7cm,height=6cm]{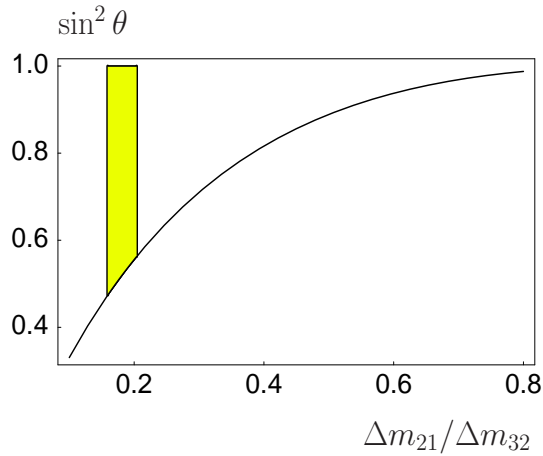}
  \caption{Allowed range of the misalignment between the vectors
    $\mathbf{h_2}$ and $\mathbf{h_3}$.}
  \label{fig:misalignment}
\end{figure}

There is a minimum value of $\sin^2\theta$ consistent with the experimentally
measured values of $R$. This value is determined by
\begin{equation}
  \label{eq:minimum}
  \sin^2\theta|_{\mbox{\tiny{min}}} =
  \left .\frac{(1 + h_r \,r_N)^2}{4 h_r\, r_N}\right|_{\mbox{\tiny{min}}}
  \left[
    1 - \left(\frac{1-R}{1+R}\right)^2
  \right]_{\mbox{\tiny{min}}}= 
  1 - 
  \left(
    \frac{1 - R_{\mbox{\tiny{min}}}}{1 + R_{\mbox{\tiny{min}}}}
  \right)^2\, ,
\end{equation}
and corresponds to the minimum misalignment between $\mathbf{h_2}$ and 
$\mathbf{h_3}$. Thus, in order to reproduce the correct solar and 
atmospheric mass scale ratio $\sin^2\theta\gtrsim 0.47$. 
Figure~\ref{fig:misalignment} shows the misalignment allowed region.

Although not consistent with neutrino experimental data there is an
interesting limit when the contribution from $N_2$ to the neutrino
mass matrix is small in comparison with those from $N_3$. In this
case the neutrino mass matrix becomes projective and therefore it can
be diagonalized with only two rotations. The rotation angles can be
written as
\begin{align}
  \label{eq:rotangles}
  \tan\theta_{23}&=-\frac{h_{23}}{h_{33}}\,,\nonumber\\
  \tan\theta_{13}&=-\frac{h_{13}}{\sqrt{h_{23}^2+h_{33}^2}}\,.
\end{align}
As it will be shown in section~\ref{sec:coll-sig} these results 
are good approximations in the case we are considering.
\section{Fermionic dark matter}
\label{sec:scalarDM}
Before discussing possible collider signals of the charged scalar
\footnote{We will denote the $\eta^\pm$ mass by $m_\eta$} we will
study the implications of the assumed mass spectrum, $M_1 \ll M_2<M_3<m_\eta$, 
on DM. In ref.~\cite{Kubo:2006yx}, $N_1$ was assumed to
be a cold DM particle. Based on this assumption, two crucial
observations, related with $m_\eta$ and the Yukawa couplings
$h_{\alpha 1}$, were made:
\begin{itemize}
\item[(i)] The following relation has to be satisfied in order to obtain the
observed DM relic density, $\Omega_d h^2 \simeq 0.12$ \cite{Spergel:2006hy}:
\begin{eqnarray}
\label{constraint1} 
\left[
\sum_{\alpha,\beta}|h_{\alpha 1} h_{\beta 1}^*|^2\right]^{1/2} & \gtrsim
& 0.08 \left( \frac{m_\eta}{100~\mbox{GeV}}\right)^2.
\end{eqnarray}
Restricting the Yukawa couplings to the perturbative
regime, i.e.  the left-hand side of (\ref{constraint1}) $\lesssim 1$,
it was found that $m_\eta\lesssim 350$ GeV.  Furthermore, the constraint
(\ref{constraint1}), being a lower bound for the Yukawa couplings
$h_{\alpha 1}$, should be compared with the constraint derived from
$\mu\to e \gamma$, which gives an upper bound for the Yukawa couplings
(see section~\ref{sec:fvclepdecays}, eq.~(\ref{eq:boundYuk})).  The
apparent contradiction between these bounds was overcome by
assuming a specific structure for the Yukawa couplings in reference
\cite{Kubo:2006yx}.
\item[(ii)] The constraint $M_1$ $\gtrsim 10$
GeV for $m_\eta \gtrsim 100$ GeV must be satisfied in addition to
the requirement that $M_1 < m_\eta$.
\end{itemize}
If (i) and (ii) are combined, the hierarchical mass relation 
$M_1/M_2<O(10^{-2})$ imply that $M_{2,3} > m_\eta$ which is not
consistent with the analysis of neutrino masses discussed in the
previous section.  Moreover, this relation, in turn, requires another
suppression mechanism for $\mu\to e \gamma$ \footnote{A suppression
mechanism based on a low-energy flavor symmetry in the same type of
models, with a radiative neutrino mass generation, was proposed in
\cite{Kajiyama:2006ww}.}.  Therefore, the assumed mass spectrum, 
$M_1 \ll M_2<M_3<m_\eta$, does not fit within the cold DM
scenario of \cite{Kubo:2006yx}. 

In what follows we will discuss whether $N_1$ can be a viable WDM
candidate. In this case there are important differences compared with
the conventional sterile neutrino WDM scenario in which sterile
neutrinos are produced by non-resonant active-sterile neutrino
oscillations \cite{Dodelson:1993je,Asaka:2006nq}, namely:
\begin{itemize}
\item[$(a)$] The decay of $N_1$ is forbidden by the $Z_2$ symmetry
\footnote{A possible origin of this symmetry was discussed in
\cite{Kubo:2006rm}.}. Thus, the X-ray
constraint~\cite{Dolgov:2000ew,Abazajian:2001vt,Mapelli:2006ej,Boyarsky:2005us},
derived from the absence of detection of X-ray photons from sterile
neutrino radiative decays, can not be applied.  This constraint, when
applied to the conventional sterile neutrino WDM scenario, yields an
upper bound of $m_{\mbox{\tiny{WDM}}}\lesssim$ 4 keV 
\cite{Boyarsky:2005us}.  This result combined with the Lyman-alpha
forest data, which lead to a lower limit of
$m_{\mbox{\tiny{WDM}}}\gtrsim$ 10-14 keV, has ruled 
out the
possibility  \footnote{If the sterile neutrinos as WDM are generated in decays
of some heavier particles, then the situation may 
change \cite{Kusenko:2006rh}.} that all the DM consists of sterile neutrinos
\cite{Seljak:2006qw,Viel:2006kd}
 (see also \cite{Gorbunov:2008ka}).
\item[$(b)$] In the conventional scenario the Yukawa couplings of the
right-handed neutrino are tiny.  Actually they can not be thermalized
without mixing with the active neutrinos \cite{Dodelson:1993je} and
therefore can not be regarded as thermal relics.  In contrast to the
conventional case, the Yukawa couplings $h_{\alpha k}$ in the current
model are not necessarily small (see section
\ref{sec:fvclepdecays}). Thus, $N_1$ can be in thermal equilibrium at
high temperatures.  This implies that the constraints discussed in the
literature on thermal WDM particles
\cite{Hannestad:2003ye,Viel:2005qj,Seljak:2006qw,Viel:2006kd} can be
applied in our case.  Of course, the largest value of $h_{\alpha k}$
must be consistent with the upper bound derived from $\mu\to e \gamma$
(see eq.~(\ref{eq:boundYuk})).
\end{itemize}
Current cosmological data constraints
\cite{Hannestad:2003ye,Viel:2005qj,Seljak:2006qw,Viel:2006kd} imply
that DM can consists of only $N_1$ if the relativistic degrees of
freedom at the decoupling temperature ($g_*(T_D)$) are larger than
$10^3$, for $M_1\lesssim 1$ keV \cite{Viel:2005qj}.  This is not
satisfied in this model, the relativistic degrees of freedom can be at
most $116$. Therefore, $N_1$ can be regarded as WDM if there exists,
in addition to $N_1$, a dominant cold DM relic that gives a
contribution of $\sim$ 90\% to the total DM relic density and if
$M_1\lesssim 16$ eV~\cite{Viel:2005qj} (this possibility, within the
conventional WDM sterile neutrino scenario, has been throughout
studied in \cite{Palazzo:2007gz}).

From a more detailed analysis of this scenario we have found that the
annihilation rate of $N_1$ at temperature $T$ can be written as
\begin{eqnarray}
\Gamma[T] &\simeq &
\left(\frac{7}{120}\right)^2 \frac{\pi^5}{54 \zeta(3)}T^5 
\,\frac{y_1^2}{m_\eta^4}~,\quad
y_1^4
\equiv\sum_{\alpha,\beta }|h_{\alpha 1} h_{\beta 1}^*|^2\,.
\end{eqnarray}
Here we have assumed $m_\eta \gg T \gg M_1$.  The decoupling
temperature can be estimated by equating the annihilation rate with
the expansion rate, $H=1.66~\sqrt{g_*(T)}~T^2/m_{pl}$. From
$H(T_D)\simeq \Gamma(T_D)$ we get
\begin{eqnarray}
\label{tdec}
y_1 \left(\frac{100\mbox{GeV}}{m_\eta}
\right)^{2}&\simeq &3.73\times 10^{-5} 
\left( \frac{g_*(T_D)}{g_*(T_\nu)} \right)^{1/4} 
\left( \frac{\mbox{GeV}}{T_D} \right)^{3/2},
\end{eqnarray}
where $T_\nu$ is the decoupling temperature of the active neutrinos
and $g_*(T_\nu)=10.75$. For $T_D\simeq 2$ GeV, for which
$g_*(T_D)=77.5$
\footnote{We have assumed that at $T_D$ only $N_1$, among $N_k$'s, remains
relativistic.}, eq.~(\ref{tdec}) becomes
\begin{equation}
\label{eq:decTemp-bound}
y_1 \left(\frac{100\mbox{GeV}}{m_\eta}
\right)^{2}\simeq 2.2\times 10^{-5}\,,
\end{equation} 
which, as we can see from eq. (\ref{eq:boundYuk}), satisfies the
constraint coming from $\mu\to e \gamma$. Note that a stringent
experimental upper limit on $Br(\mu\to e\gamma)$ will imply a larger
decoupling temperature. For example, a three orders of magnitude more
stringent bound on $Br(\mu\to e\gamma)$, as the one expected in near
future experiments~\cite{meg}, will enforce $T_D$ to be larger than
$\sim140$ GeV.
\section{Collider physics}
\label{sec:coll-sig}
The Yukawa couplings that govern neutrino physics also determine the
fermionic two-body decays of $\eta_{R,I}^0$ and $\eta^\pm$. According
to the Yukawa interactions in (\ref{eq:yukint}) possible decays of
these states are:
\begin{align}
  \label{eq:neutral-Sca-decay}
  \eta_{R,I}^0&\to \nu_\alpha\,N_i\\
  \label{eq:charged-Sca-decay}
  \eta^\pm&\to \ell_\alpha^\pm\,N_i\,.
\end{align}
As will be discussed below $N_{2,3}$ follow decay chains that can lead
to only missing energy. In that case the observation of the
neutral Higgses $\eta^0_{R,I}$ will be problematic. On the contrary,
since charged scalar final states always contain --at least-- a
charged lepton their decays are easier to observe. Therefore, we will
focus on charged Higgs decays. Apart from the Yukawa interactions the
scalar doublet $\eta$ has also gauge (and scalar) interactions which
induce the decays $\eta^\pm\to \eta_{R,I}^0\,W^\pm$, if kinematically
possible.

At LHC charged scalars can be produced either in association with a
neutral scalar (single production) or in pairs \cite{Djouadi:2005gj}.
In the former case the mechanism proceeds via $q\overline{q}$
annihilation mediated by a virtual $W$ vector boson whereas in the
latter case through s-channel exchange of a virtual $\gamma$ and a
$Z$:
\begin{align}
  \label{eq:prod-mechanism}
  q\overline{q} &\to \eta^\pm\eta_{R,I}^0\\
  q\overline{q} &\to\eta^+\eta^-\,.
\end{align}
Charged scalar production in association with an $\eta_I^0$ has been
calculated in reference \cite{Cao:2007rm}.  According to this result
the production cross section is larger than 100 fb for $m_\eta\lesssim
200\;$GeV.  The pair production cross section, on the other hand, is
further suppressed as it can not exceed 10 fb for charged scalar
masses below 250 GeV~\cite{Djouadi:2005gj}. Contrary, at ILC the pair
production cross section is larger than 10 fb for $m_\eta\lesssim 350\;$GeV
\cite{Battaglia:2001be}. Thus, depending on the accumulated
luminosity, LHC (ILC) should be able to explore up to masses of order
$m_\eta\sim 200-250\;$GeV (400 GeV).

\subsection{Right-handed neutrinos: Decays, signals and identification}
\label{sec:decays-signals-ident}
The correlations between charged scalar decays and neutrino mixing
angles which will be discussed latter could be studied in collider
experiments only if the decaying right-handed neutrino can be
identified. Experimentally, in principle, this can be done. Let us
discuss this in more detail: right-handed neutrinos, stemming from
charged scalar decays, will produce, via an off-shell $\eta^\pm$,
charged leptons through the decay chains
\begin{align}
  \label{eq:decay-chains1}
  N_3    &\to \ell^\pm_\alpha\eta^\mp \to 
  \ell^\pm_\alpha \ell^\mp_\beta N_2 \to 
  \ell^\pm_\alpha \ell^\mp_\beta \ell^\pm_{\alpha'}\eta^\mp \to 
  \ell^\pm_\alpha \ell^\mp_\beta \ell^\pm_{\alpha'}\ell^\mp_{\beta'} N_1\\
  \label{eq:decay-chains2}
  N_{3,2} &\to \ell^\pm_\alpha\eta^\mp \to \ell^\pm_\alpha\ell^\mp_\beta N_1\,.
\end{align}
In addition to these decay chains there are others which involve
neutral scalars and lead to either {\it dilepton + missing energy}
($\ell_\alpha^\pm\ell_\beta^\mp\nu_{\alpha'}\nu_{\beta'} N_1$) or {\it
  missing energy} ($\nu_\alpha\nu_\beta\nu_{\alpha'}\nu_{\beta'} N_1$
or $\nu_\alpha\nu_\beta N_1$) signals.

The most important signatures for the identification procedure are
(\ref{eq:decay-chains1}) and (\ref{eq:decay-chains2}) due to their low
backgrounds \cite{deCampos:2007bn}. The right handed neutrino
identification from the remaining decay chains might be rather hard as
they involve additional missing energy.  Thus, in general, they will
diminish the relevant signals. Whether the decay branching ratios for
the processes in (\ref{eq:decay-chains1}) and (\ref{eq:decay-chains2})
can dominate depend upon the different parameters (mainly Yukawa
couplings and scalar masses), which we will now discuss in turn. The
decay chains in eq. (\ref{eq:decay-chains1}) dominates over the
processes $N_3\to \ell^\pm_\alpha \ell^\mp_\beta
\nu_{\alpha'}\nu_{\beta'} N_1$ and $N_3\to \nu_\alpha \nu_\beta
\nu_{\alpha'}\nu_{\beta'} N_1$ as long as
\begin{equation}
  \label{eq:condition1}
  \sum_{\substack{\alpha,\beta\\
      \alpha',\beta'}}
  Br(N_3\to\ell^\pm_\alpha \ell^\mp_\beta \ell^\pm_{\alpha'}\ell^\mp_{\beta'} N_1)
  >
  \begin{cases}
    \sum_{\substack{\alpha,\beta\\
        \alpha',\beta'}} 
    Br(N_3\to\ell^\pm_\alpha \ell^\mp_\beta \nu_{\alpha'}\nu_{\beta'} N_1)\\
    \sum_{\substack{\alpha,\beta\\
        \alpha',\beta'}}
    Br(N_3\to\nu_\alpha \nu_\beta \nu_{\alpha'}\nu_{\beta'} N_1)\,.
  \end{cases}
\end{equation}
The conditions on the parameter space of the model for which
\eqref{eq:condition1} is fulfilled can be entirely determined from
the three-body decay processes $N_i\to \ell_\alpha^\pm\ell_\beta^\mp
N_j$ and $N_i\to \nu_\alpha\nu_\beta N_j$ as the branching ratios in
\eqref{eq:condition1} are given by
\begin{equation}
  \label{eq:prod-BR}
  \sum_{\substack{\alpha,\beta\\
      \alpha',\beta'}}
    Br(N_3\to f_\alpha f_\beta f'_{\alpha'}f'_{\beta'} N_1)
    =\sum_{\substack{\alpha,\beta\\
      \alpha',\beta'}}
    Br(N_3\to f_\alpha f_\beta N_2)\times 
    Br(N_2\to f'_{\alpha'} f'_{\beta'} N_1).
\end{equation}
Thus, from eq. \eqref{eq:prod-BR} and using the shorthand notation
\begin{eqnarray}
  \label{eq:shothand-N}
  Br(N_i\to N_j) &=&  \sum_{\alpha,\beta}Br(N_i\to \ell_\alpha^\pm \ell_\beta^\mp N_j)\,,\\
  Br_{\text{inv}}(N_i\to N_j) &=&  \sum_{\alpha,\beta}Br(N_i\to \nu_\alpha \nu_\beta N_j)\,,
\end{eqnarray}
the constraints in \eqref{eq:condition1} become
\begin{align}
  \label{eq:constraints2}
  Br(N_2\to N_1) >& Br_{\text{inv}}(N_2\to N_1)\nonumber\\
  Br(N_3\to N_2) \times Br(N_2\to N_1) >&
  Br_{\text{inv}}(N_3\to N_2) \times Br_{\text{inv}}(N_2\to N_1)\,.
\end{align}
Similar conditions can be also obtained in the case of the decay
chains in (\ref{eq:decay-chains2}), namely
\begin{equation}
  \label{eq:condition3}
  Br(N_i\to N_j) > Br_{\text{inv}}(N_i\to N_j)\,.
\end{equation}

The partial decay width for the process $N_i\to f_\alpha f_\beta N_j$ summed over
all possible final states for a fixed $j$, is given by
\begin{equation}
  \label{eq:partial-decay-width}
  \sum_{\alpha,\beta}\Gamma(N_i\to f_\alpha f_\beta N_j) = 
  \frac{|\mathbf{h}_i|^2 |\mathbf{h}_j|^2+ (\mathbf{h}_i\cdot \mathbf{h}_j)^2}{384\pi^3}
  \frac{M_j^5}{m_S^4} \,I(M_i/M_j)
\end{equation}
where
\begin{equation}
  \label{eq:loop-int}
  I(x) = 1 - 8x^2 - 24x^4\,\ln(x) + 8x^6 - x^8\,
\end{equation}
and $S=\eta$ if $f=\ell$ or $S=\eta_{R,I}$ if $f=\nu$. This expression,
in addition to the conditions \eqref{eq:constraints2} and
\eqref{eq:condition3}, lead to the constraint
\begin{equation}
  \label{eq:mR_meta_constraint}
  m_{\eta_{R,I}}>m_\eta\,.
\end{equation}
Consequently, as long as the neutral scalars become heavier than the
charged one the decay processes in (\ref{eq:decay-chains1}) and
(\ref{eq:decay-chains2}) become dominant. Note that this result holds
only if $N_{2,3}$ decay inside the detector. Whether this is indeed
the case depends on the parameters that define
eq.~\eqref{eq:partial-decay-width}. Since right-handed neutrino masses
$M_{2,3}$ as well as the parameter space vectors $|\mathbf{h_{2,3}}|$
are bounded by neutrino physics, once the constraint
\eqref{eq:mR_meta_constraint} is imposed \footnote{Scalar masses are
  also constrained from the requirement of scalar production at LHC or
  ILC (upper bound) and from LEP data (lower bound).} the only free
parameter is $\mathbf{h_1}$. Accordingly, the right-handed neutrino
decay lengths are strongly determined by the value of
$|\mathbf{h_1}|$. We calculate $N_2$ and $N_3$ decay lengths by
randomly varying the Yukawa couplings $h_{\alpha i}$ for the benchmark
point $m_{R,I}=140 \,\text{GeV}$, $m_\eta=150 \,\text{GeV}$, $M_{2}=25
\,\text{GeV}$ and $M_{3}=45 \,\text{GeV}$. After imposing neutrino
physics contraints at 1$\sigma$ level \cite{Maltoni:2004ei} we get
\begin{equation}
  \label{eq:decay-length}
  L_2 \subset [0.08, 300] \,\text{m},\quad L_3 \subset [10^{-3}, 2] \,\text{m}\,,
\end{equation}
which shows that $N_3$ always decay within the detector whereas $N_2$
decays might occur outside. 

As can be seen from eq.~\eqref{eq:partial-decay-width} the larger
(smaller) $|\mathbf{h_1}|$ the smaller (larger) $L_2$. For the
benchmark point we have considered, it has been found that in those
regions of parameter space in which $L_2$ is smaller than few meters
$Br(N_3\to N_2)\sim{\cal O}(10^{-2})$ which implies that most $N_3$
decays will proceed through the decay chains in
(\ref{eq:decay-chains2}).  On the contrary, when $L_2$ is large $N_2$
will behave, from the collider point of view, as $N_1$ and the only
possible signals will be either {\it dilepton + missing energy} or
{\it missing energy}. In this case according to our results the
process $N_3\to \ell_\alpha^\pm\ell_\beta^\mp N_1$ will be sizable
($Br(N_3\to N_1)>0.1$).

In general, since from eq. \eqref{eq:partial-decay-width} we have
\begin{equation}
  \label{eq:N2toN1}
  Br(N_2\to N_1) = \frac{m_{R,I}^4}{m_{R,I}^4 + m_\eta},
\end{equation}
if $m_\eta\ll m_{R,I}$ small values of $|\mathbf{h_1}|$ will enhance
the decays in \eqref{eq:decay-chains1}. For the smallest value of
$|\mathbf{h_1}|$ for which $N_2$ still decays inside the detector
(typically $10^{-3}$) we found that
\begin{equation}
  \label{eq:upperboundN3toN2toN1}
  Br(N_3\to N_2)\times Br(N_2\to N_1)\lesssim 0.5\,.
\end{equation}

Hard leptons with missing energy (eqs. (\ref{eq:decay-chains1}) and
(\ref{eq:decay-chains2})) are typical accelerator signatures in
conserving and non-conserving R-parity violating supersymmetric models
\cite{Barnett:1993ea,deCampos:2007bn}. Indeed, as pointed out in
references \cite{Barnett:1993ea,deCampos:2007bn}, the discovery of
supersymmetry could arise from such a signal. In the present case the
possibility of having in addition displaced vertices might facilitate
the reconstruction of $N_2$ and $N_3$.  Actually, since $W$ and $Z$
leptonic decay modes occur at the interaction point, this type of
signals are practically background free once the dipleton invariant
mass distribution from the displaced vertex is above 10 GeV
\cite{deCampos:2007bn}.

Regarding the identification procedure if $N_3$ decay according to
(\ref{eq:decay-chains1}) the identification might be possible by
counting the number of leptons emerging from a given vertex. In
contrast to the decay chain (\ref{eq:decay-chains1}), if $N_3$ follows
the processes in (\ref{eq:decay-chains2}) the number of leptons from
$N_{3,2}$ decays will be the same, and the charged lepton counting
``method'' can not be used. In this case $N_3$ from $N_2$ decays can
be distinguished by looking to the kinematic endpoint of the lepton
pair invariant mass distribution. This method have been extensively
discussed in the MSSM context~\cite{Allanach:2000kt} and might be also
applicable in this case. Note that the kinematic endpoint technique
could be also applicable when $N_3$ follows the decay chain
(\ref{eq:decay-chains1}).  Thus, the right-handed neutrino
identification procedure can be entirely based on this method.

\subsection{Collider signals related to neutrino physics}
\label{sec:collider-signals-neutrino-pp}
The results presented below were obtained by numerically diagonalizing
eq. (\ref{eq:nmmappx}) for random parameters and checking for
consistency with experimental neutrino
constraints~\cite{Maltoni:2004ei}.  Different correlations among
neutrino mixing angles and charged scalar decay branching ratios were
found as expected from eq. (\ref{eq:rotangles}). The parameter $m_0$,
which essentially corresponds to $m_R$ or $m_I$, was taken in the
range $100\;\mbox{GeV}\leq m_0\leq 400\;\mbox{GeV}$ \footnote{The
  charged scalar mass was also taken in this range.} whereas the
masses of $N_3$ and $N_2$ between $40\;\mbox{GeV}\leq M_3\leq
50\;\mbox{GeV}$ and $20\;\mbox{GeV}\leq M_2\leq 30\;\mbox{GeV}$
\footnote{$M_1$ was taken below 16 eV as required by DM constraints}.
The Yukawa couplings were chosen such that
$|\mathbf{h_2}|/|\mathbf{h_3}|\subset [0.4,0.9]$. In regions of
parameter space in which $N_3$ and $N_2$ are comparable --though
lighter-- to $m_\eta$ the correlations, discussed below, are less
pronounced. However, the decays chains (see eqs.
(\ref{eq:decay-chains1}) and (\ref{eq:decay-chains2})) will involve
hard leptons from which the right-handed neutrinos can be readily
identified.  On the other hand, if $N_3$ and $N_2$ are much more
lighter than $\eta^\pm$ the data points become strongly correlated. In
this case, in contrast to the previous one, charged leptons emerging
from the decay chains might be near the $\tau$ --and possibly $\mu$--
threshold which will render the right-handed neutrino identification
problematic.

\begin{figure}[t]
  \centering
  \includegraphics[height=8cm,width=9cm]{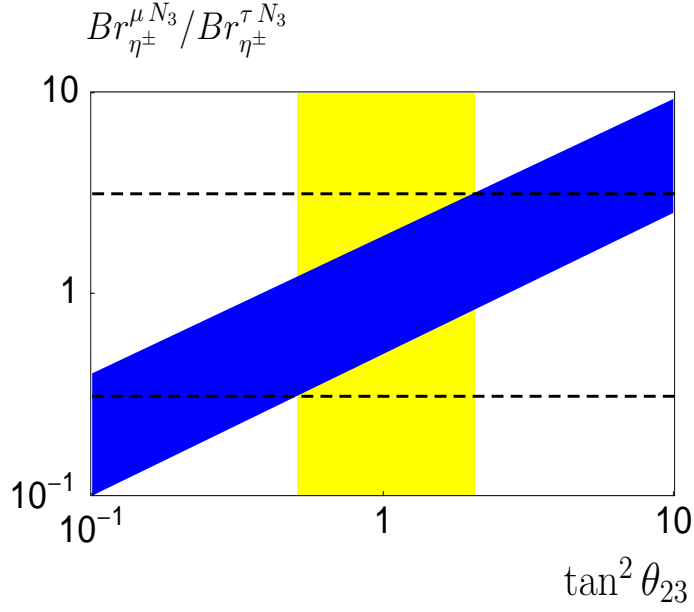}
  \caption{Ratio of decay branching ratios
  $Br^{\mu\,N_3}_{\eta^\pm}/Br^{\tau\,N_3}_{\eta^\pm}$ versus
  $\tan^2\theta_{23}$. The vertical strip indicates the current $3\sigma$ range
  for $\tan^2\theta_{23}$ whereas the horizontal lines indicate the predicted
  range for this observable.}
\label{fig:atmangle-corr}
\end{figure}
Figure \ref{fig:atmangle-corr} shows that the ratio of decay branching
ratios $Br^{\mu\,N_3}_{\eta^\pm}/Br^{\tau\,N_3}_{\eta^\pm}$
($Br(\eta^\pm \to \ell^\pm\,N_k)\equiv Br^{\ell\,N_k}_{\eta^\pm}$) is
correlated with $\tan^2\theta_{23}$. From the best fit point value
($\tan^2\theta_{23}=1$) $Br^{\mu\,N_3}_{\eta^\pm}\simeq
Br^{\tau\,N_3}_{\eta^\pm}$ is expected.  Furthermore, the 3$\sigma$
range for the atmospheric mixing angle allows to predict this
observable to lie within the interval [0.35,3.0], as indicated by the
horizontal dashed lines in figure~\ref{fig:atmangle-corr}.

\begin{figure}[t]
  \centering
  \includegraphics[height=8cm,width=9cm]{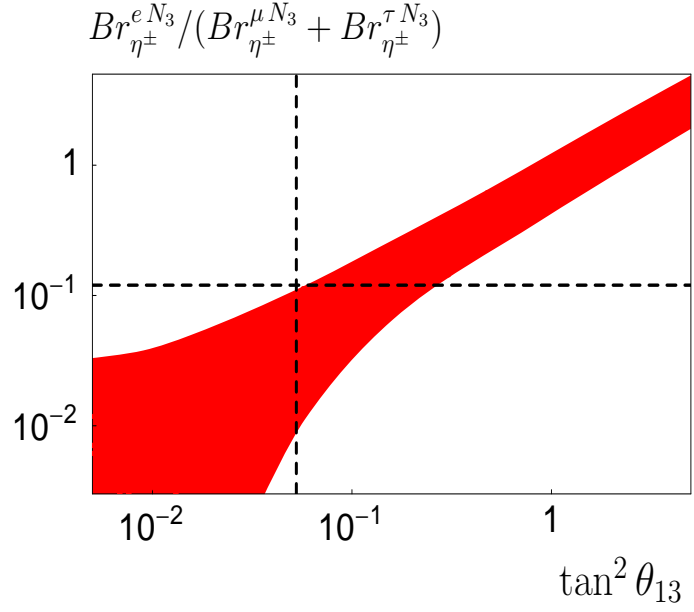}
  \caption{Ratio of decay branching ratios
  $Br^{e\,N_3}_{\eta^\pm}/(Br^{\mu\,N_3}_{\eta^\pm}+ Br^{\tau\,N_3}_{\eta^\pm})$ versus
  $\tan^2\theta_{13}$. The vertical line indicates the current $3\sigma$ upper
  bound for $\tan^2\theta_{13}$ whereas the horizontal lines indicate the predicted
  upper bound for this observable.}
  \label{fig:reactorangle-corr}
\end{figure}
We have found that there exits an upper bound on the ratio of decay
branching ratios
\begin{equation}
  \label{eq:electronFS}
  \frac{Br^{e\,N_3}_{\eta^\pm}}{Br^{\mu\,N_3}_{\eta^\pm} + 
    Br^{\tau\,N_3}_{\eta^\pm}}
  \lesssim 1.2\times 10^{-1}
\end{equation}
derived from the correlation between this observable and
$\tan^2\theta_{13}$ and demonstrated by
figure~\ref{fig:reactorangle-corr}. From this bound
$Br^{e\,N_3}_{\eta^\pm}$ is expected to be around 1 order of
magnitude smaller than $Br^{\mu\,N_3}_{\eta^\pm} +
Br^{\tau\,N_3}_{\eta^\pm}$. Which in turn implies, according to
$Br^{\mu\,N_3}_{\eta^\pm}\simeq Br^{\tau\,N_3}_{\eta^\pm}$, that
$e\,N_3$ final states are further suppressed than $\mu\,N_3$ and
$\tau\,N_3$ final states.

\begin{figure}[t]
  \centering
  \includegraphics[height=8cm,width=9cm]{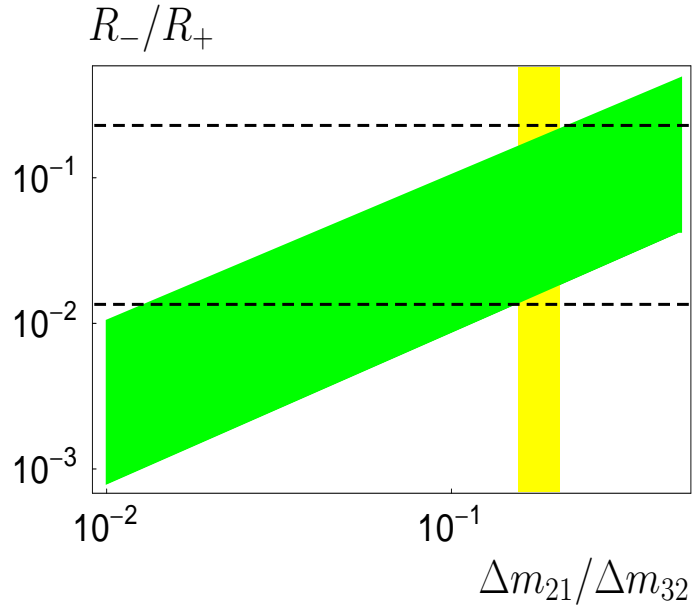}
  \caption{Ratio of decay branching ratios $R_-/R_+$ versus 
    $\Delta m_{12}/\Delta m_{23}$. The vertical
    shaded strip indicate the current 3$\sigma$ range for 
    $\Delta m_{12}/\Delta m_{23}$ whereas the
    horizontal dashed lines shows the allowed region for $R_-/R_+$.}
  \label{fig:rat-brcom}
\end{figure}

From eqs.~(\ref{eq:fermionicDMeig}) and (\ref{eq:atmospsolarR}) we
found a quantity, $R_-/R_+$, which is related to
$R$. $R_{\mp}$ can be written as
\begin{equation}
  \label{eq:decaybrat-com}
  \frac{R_{\mp}}{{\cal  R}_f} = 
  1 \mp 
  \left[
    1 - 4
    r_N\frac{
      \sum_{i,j}Br_\eta^{\ell_i N_2}Br_\eta^{\ell_j N_3}
      - \left(
        \sum_i \sqrt{Br_\eta^{\ell_i N_2} Br_\eta^{\ell_i N_3}}
      \right)^2}
    {\left(
        r_N\sum_i Br_\eta^{\ell_i N_2} + \sum_i
        Br_\eta^{\ell_i N_3}
      \right)^2}
  \right]^{1/2}\,,
\end{equation}
where $i,j$ run over $e, \mu, \tau$, $r_N$ corresponds to the
right-handed neutrino mass ratio defined in eq.~(\ref{eq:r-h-n-ratio})
and ${\cal R}_f$ is a common global factor that involves the same
parameters that define ${\cal G}_f$ (see eq.~(\ref{eq:globalfactor}))
and decay branching ratios. Note that in the ratio $R_-/R_+$ this
factor cancel. Numerical results are shown in
figure~\ref{fig:rat-brcom}. The spread in the plot is due to an
ambiguity in the sign of the Yukawa couplings. From the current
3$\sigma$ range for $\Delta m_{12}/\Delta m_{23}$ (vertical shaded
strip in figure \ref{fig:rat-brcom}) this quantity is predicted to
lie in the range (horizontal dashed lines) $[1.4\times
10^{-2},2.0\times 10^{-1}]$.

As long as the constraints $|\mathbf{h_2}|/|\mathbf{h_3}|< 1$ and
$M_2/M_3<1$ are satisfied the contributions of $N_2$ to the neutrino
mass matrix are small in comparison with those from $N_3$.  Thus, the
atmospheric and reactor angles approximate to
eqs.~(\ref{eq:rotangles}). The results shown in figs.
\ref{fig:atmangle-corr} and \ref{fig:reactorangle-corr} can be
understood as a consequence of these constraints. Note that the
sharpest correlations among the decay patterns of the charged scalar
with neutrino mixing angles are obtained for the largest allowed (by
neutrino experimental data) hierarchies between the parameter space
vectors $|\mathbf{h_2}|$ and $|\mathbf{h_3}|$ and the right-handed
neutrino masses $M_2$ and $M_3$. 

In order to generate an inverted light neutrino mass spectrum $({\cal
  M}_\nu)_{11}$ has to be large (of the same order of $({\cal
  M_\nu})_{22,33,23}$). Thus, large contributions from the loop
involving $N_2$ are necessary.  These contributions {\it spoil} the
leading projective nature of the neutrino mass matrix and therefore
the atmospheric and reactor angles are no longer determined by
eqs.~(\ref{eq:rotangles}).  Accordingly, the correlations among
collider observables and neutrino mixing angles we have discussed will
not hold in this case. However, in principle, these results can be
used to discriminate between the normal and inverted mass hierarchies
as follows: If $M_3>M_2$ and $\sum_\alpha Br_\eta^{\ell_\alpha
  N_2}/\sum_\alpha Br_\eta^{\ell_\alpha N_3}<1$ \footnote{This
  relation is derived from the constraint
  $|\mathbf{h_2}|/|\mathbf{h_3}|<1$} are experimentally established
but none of the observables given in figs. \ref{fig:atmangle-corr},
\ref{fig:reactorangle-corr} are found to be in the range predicted by
neutrino physics the normal mass spectrum will be excluded.

\begin{figure}[t]
  \centering
  \includegraphics[height=8cm,width=9cm]{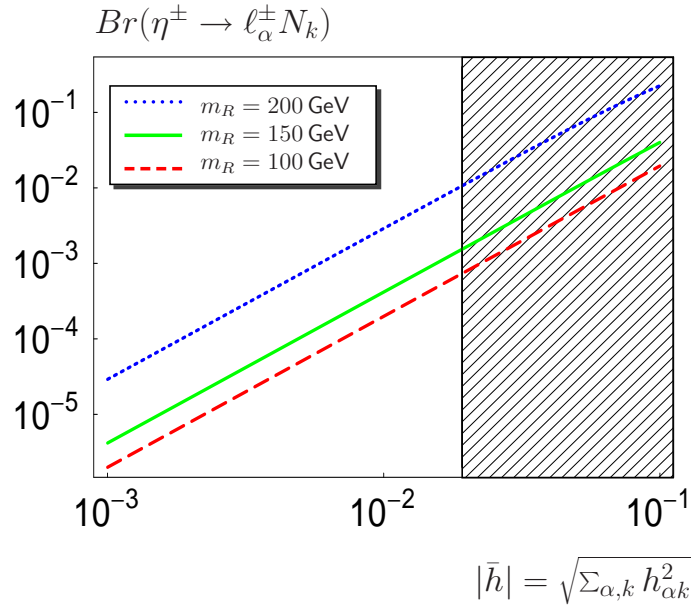}
  \caption{Charged scalar fermionic decay branching ratio as a function of the
    average Yukawa coupling $|\bar{h}|$ for the case in which $W$ gauge boson
    final states are kinematically open. The charged scalar mass has been
    fixed to 300 GeV. The shaded region is excluded by the
    experimental upper bound on $Br(\mu\to e\gamma)$.}
  \label{fig:gaugebosonsfs}
\end{figure}

$W^\pm$ final states are also possible depending on whether the mass
difference $\Delta M = m_\eta-m_{R,I}$ is larger than $M_W$ . Once
kinematically open, this decay channel will dominate over the
fermionic final states. However, as illustrated in
figure~\ref{fig:gaugebosonsfs}, even in that case the fermionic decay
branching ratios can be as large as $\sim 10^{-2}$. Albeit possibly
problematic to be measured at LHC might be measurable at ILC. As
indicated in figure~\ref{fig:gaugebosonsfs} (shaded region) larger
values of these branching ratios are excluded by the current upper
bound on $Br(\mu\to e\gamma)$ (see next section).
 
\section{Flavour violating charged lepton decays}
\label{sec:fvclepdecays}
In this section we will derive upper bounds on the Yukawa couplings,
$h_{\alpha k}$, and briefly discuss possible low energy lepton flavour
violating signals of the model.  The set of Yukawa interactions
induced by the right-handed neutrinos and the $SU(2)$ doublet $\eta$
are responsible for lepton flavour violating radiative decays of the
type $\l_\alpha\to l_\beta\gamma$. Here we will concentrate on $\mu\to
e\gamma$. The bounds derived from $\tau\to e\gamma$ and
$\tau\to\mu\gamma$ decays are much more weaker than those from $\mu\to
e\gamma$ and thus we will not consider them.

In the limit $m_\beta\ll m_\alpha$ the partial decay width of
$\l_\alpha\to l_\beta\gamma$, induced by $\eta^\pm$ and $N_k$, can be
written as
\footnote{Note that we are considering the case $M_k^2\ll m_\eta^2$
  and therefore the loop function reduces to a factor of 1/6.}
\begin{equation}
  \label{eq:ljtoligamma}
  \Gamma(\l_\alpha\to l_\beta\gamma) = 2\alpha m_\alpha^3
  \left(
    \frac{m_\alpha}{96\pi^2}
  \right)^2 
  \frac{\left|\sum_{k=1}^3 h_{\alpha k}^* h_{\beta k}\right|^2}{m_\eta^4}\,.
\end{equation}
From the above expression the decay branching ratio for $\mu\to
e\gamma$ can be written as
\begin{equation}
  \label{eq:brmutoegamma}
  Br(\mu\to e\gamma) \simeq \frac{\Gamma(\mu\to e\gamma)}
  {\Gamma(\mu\to e\bar{\nu}_e \nu_\mu)} = \frac{\alpha}{24\pi\,G_F^2}
  \,\frac{\left|\sum_{k=1}^3 h_{1 k}^* h_{2 k}\right|^2}{m_\eta^4}\,,
\end{equation}
and the current upper bound on this process yields the upper bound
\begin{equation}
  \label{eq:yukawasandmeta}
  \left|\sum_{k=1}^3 h_{1 k}^* h_{2 k}\right|\lesssim 4.1\times 10^{-5}
  \left(\frac{m_\eta}{100\,\mbox{GeV}}\right)^2\,.
\end{equation}
The largest value for these Yukawa couplings is derived from the
largest charged scalar mass, $m_\eta$ = 400 GeV, in this case
\begin{equation}
  \label{eq:boundYuk}
  \left|\sum_{k=1}^3 h_{1 k}^* h_{2 k} \right|\lesssim 6.5\times 10^{-4}\,.
\end{equation}
For smaller charged scalar masses the bound becomes more
stringent. Note that this constraint can be satisfied by either
$h_{1k}\ll h_{2k}$ or $h_{2k}\ll h_{1k}$. An exception being the case
of non-hierarchical Yukawa couplings. Under this assumption, and for
$m_\eta=$ 300 GeV, the upper bound $h\lesssim 1.9\times 10^{-2}$ can
be placed. This constraint corresponds to the shaded region shown in
figure \ref{fig:gaugebosonsfs}.

\begin{figure}[t]
  \centering
  \includegraphics[height=8cm,width=9cm]{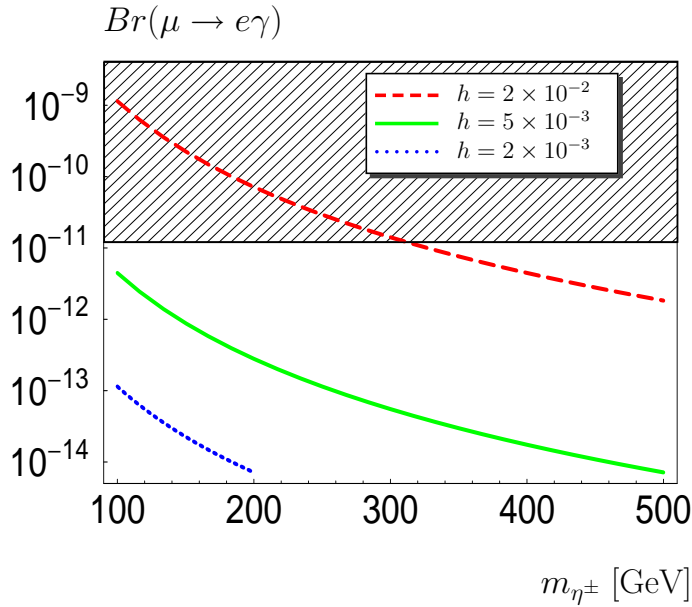}
  \caption{$Br(\mu\to e\gamma)$ as a function of the charged scalar
  mass under the assumption of non-hierarchical Yukawa couplings. The
  shaded region is excluded by the current experimental upper bound.}
  \label{fig:mutoegamma}
\end{figure}
Since we do not have a theory for the Yukawas, an absolute value for
$Br(\mu\to e\gamma)$ can not be predicted. However, assuming
non-hierarchical couplings this branching ratio is found to
be larger than $10^{-14}$ for $h\gtrsim 10^{-3}$ as shown in
figure~\ref{fig:mutoegamma}. Note that this result is a
consequence of the assumption ${\cal O}(h_{1k})\approx
{\cal O}(h_{2k})$ and not a general feature of the model.

\section{Summary}
\label{sec:summary}
Assuming the mass spectrum $M_1\ll M_2< M_3< m_\eta$ we have studied
some phenomenological aspects of the radiative seesaw model
\cite{Ma:2006km}. In particular, we showed that
current experimental neutrino data can be used to constraint the
parameter space of the model. Thus, some fermionic decays of the
charged scalar $\eta^\pm$ are correlated with neutrino mixing angles
which in turn allows to predict several ratios of decay branching
ratios. Especially interesting is that if the $\eta^\pm$ is produced
at colliders similar number of events with $\tau$ and $\mu$ final
states are expected, whereas events with $e$ are expected to be small.
As has been said, these predictions could be tested in accelerator
experiments depending on whether the decaying right-handed neutrino
can be identified. We have discussed how this could be achieved by
either counting the numbers of leptons emerging from a given vertex or
by looking to the kinematic endpoint of the lepton pair invariant mass
distribution~\cite{Allanach:2000kt}.

We have found that the lightest sterile neutrino is a WDM particle
which, though stable, can not be the only DM component of the
Universe.  Its contribution to the DM relic density is less than
10\%. Therefore, other cold DM relic must be responsible of the
remaining 90\%.

Finally we have derived upper bounds on the Yukawa couplings of the
model from the experimental upper limit on $Br(\mu\to e\gamma)$. We
have shown that under the assumption of non-hierarchical
Yukawa couplings $Br(\mu\to e\gamma)$ is found to be larger than
$10^{-14}$ for $h\gtrsim 10^{-3}$, i.e within the range of near future
experiments~\cite{meg}.

\section{Acknowledgments}
D.A.S wants to thanks E. Nardi, Carlos E. Yaguna and J. Kamenik for
useful comments. Specially to M. Hirsch for critical readings of the
manuscript, for very useful suggestions and for pointing out an error
in the first version of the paper. Work partially supported by
Colciencias in Colombia under contract 1115-333-18740. D.A.S is
supported by an INFN postdoctoral fellowship.

\end{document}